\documentclass[a4paper,preprint,showpacs,amssymb,pre,superscriptaddress]
              {revtex4}
\usepackage{graphicx}

\begin{document}

\title{System Size Stochastic Resonance from the Viewpoint of the
Nonequilibrium Potential}

\author{Horacio S. Wio}
\email{wio@ifca.unican.es} \affiliation{Instituto de F\'{\i}sica
de Cantabria, E-39005 Santander, Spain} \affiliation{Centro
At\'omico Bariloche, 8400 San Carlos de Bariloche, Argentina}

\begin{abstract}
We analyze the phenomenon of system size stochastic resonance in a
simple spatially extended system by exploiting the knowledge of
the nonequilibrium potential. We show that through the analysis of
that potential, and particularly its ``symmetry", we can obtain a
clear physical interpretation of this phenomenon in a wide class
of extended systems, and also analyze, for the same simple model,
the effect of a general class of boundary conditions (albedo) on
this kind of phenomena.
\end{abstract}


\maketitle

\normalsize

Among the many noise induced phenomena extensively studied during
the last few decades like: stochastic resonance \cite{RMP},
noise-induced transitions \cite{lefev}, noise-induced phase
transitions \cite{nipt1}, noise-induced transport
\cite{Ratch2,nipt3}, noise-induced limit cycles \cite{mangwio};
\textit{stochastic resonance} (SR) detaches as one that attracted
--and still attracts-- the attention of a large number of
researchers, probably due to its interest from a technological as
well as a biological points of view. There is a broad range of
phenomena for which this mechanism can offer an explanation as has
been put in evidence by many reviews and conference proceedings
\cite{RMP}.

The fingerprint of the SR phenomenon is the {\em enhancement\/} of
the output signal-to-noise ratio (SNR) caused by the injection of
an {\em optimal} amount of noise into a periodically --or even
aperiodically-- driven nonlinear system. Such enhancement is the
result of a cooperative effect arising from the interplay between
{\em deterministic\/} and {\em random\/} dynamics in a {\em
nonlinear\/} system. In almost all the studies of SR, the relevant
control variable of the phenomenon was the noise intensity, while
the system's size didn't play any relevant role. However, some
recent studies on biological models of the Hodgkin-Huxley type
\cite{SSSR1,SSSR2} have shown that ion concentrations along
biological cell membranes presents intrinsic SR-like phenomena as
the number of ion channels is varied. A related result
\cite{SSSR3} shows that even in the absence of external forcing,
the regularity of the collective firing of a set of coupled
excitable FitzHugh-Nagumo units results optimal for a given value
of the number of elements. From a physical system point of view,
the same phenomenon --that has been called \textit{system size
stochastic resonance} (SSSR)-- has also been found in an Ising
model as well as in a set of globally coupled units described by a
$\phi ^{4}$ theory \cite{SSSR4}. It was even shown to arise in
opinion formation models \cite{SSSR5}.

In a recent series of papers
\cite{extend2,extend2a,extend2b,extend3a,extend3b} the phenomenon
of SR in extended systems was studied exploiting the concept of
\textit{nonequilibrium potential} (NEP). This potential is a
special Lyapunov functional of the associated deterministic system
which, for nonequilibrium systems, plays a role similar to that
played by a thermodynamic potential in equilibrium thermodynamics
\cite{GR}. Such a NEP, closely related to the solution of the time
independent Fokker-Planck equation of the system, characterizes
the global properties of the dynamics, that is: attractors,
relative (or nonlinear) stability of these attractors, height of
the barriers separating attraction basins, allowing to evaluate
the transition rates among the different attractors. In
\cite{I0,I1}, this approach was applied to the global stability
analysis of some reaction-diffusion systems. A kind of
``mini-review" on the application of this approach to the study of
SR in one, two, and three species reaction-diffusion systems could
be found in \cite{extend3b}. However, in spite of its
potentiality, this kind of approach has not been exploited for the
study of SSSR.

In this paper we present an analysis of the SSSR phenomenon in a
simple spatially extended system, exploiting previous results
obtained using the notion of NEP within the context of a simple
reaction--diffusion model. The specific model we shall focus on,
with a known form of the Lyapunov function, corresponds to a
one--dimensional, one--component model \cite{WW,RL} that, with a
piecewise linear form, mimics general bistable reaction--diffusion
models \cite{WW}. In particular we will exploit some of the
results for the influence of general boundary conditions (called
``albedo") found in \cite{SW} as well as previous studies of the
NEP \cite{I0,I1} and of SR \cite{extend2,extend2a}.

The particular non-dimensional form of the model that we work with
is \cite{SW,extend2,extend2a}
\begin{equation}
\label{Ballast}
\partial_t \phi=\partial^2_{yy} \phi-\phi+\phi_h \theta (\phi-\phi_c).
\end{equation}
We consider here a class of stationary structures $\phi(y)$ in the
bounded domain $y \in [-y_L,y_L]$ with albedo boundary conditions
at both ends, $\left. {d\phi \over dy} \right|_{y=\pm y_L} = \mp k
\,\phi (\pm y_L)$, where $k>0$ is the albedo parameter. These are
the  spatially  symmetric solutions to Eq.(\ref{Ballast}) already
studied  in \cite{SW}. The explicit forms of these stationary
patterns --not shown here-- are given by Eq. (9) of \cite{SW} (see
also Fig. 4 in \cite{SW}).

The double-valued coordinate $y_c$, at which $\phi = \phi_c$, is
given by
\begin{equation}
\label{yc}
y_c^\pm = {1\over 2}y_L - {1\over 2} \ln \left[
{z\gamma(k,y_L) \pm \sqrt{z^2 \gamma(k,y_L)^2+1-k^2} \over 1+k}
\right],
\end{equation}
with $\gamma(k,y) = \sinh(y)+k\ \cosh (y)$, and
$z=1-2\phi_c/\phi_h$ ($-1< z< 1$).

When $y_c^\pm$ exists and $y_c^\pm < y_L$, this pair of solutions
represents a  structure with a  central ``excited'' zone
($\phi>\phi_c$) and two lateral ``resting'' regions
($\phi<\phi_c$). For each parameter set, there are two  stationary
solutions, given by the two  values of $y_c$. Figure 5 in
\cite{SW} depicts the curves corresponding to the relation
$y_c/y_L$ vs. $k$, for various values of $\phi_c/\phi_h$.

It has been shown \cite{SW} that the structure with the smallest
``excited'' region (with $y_c=y_c^+$, denoted by $\phi_u(y)$) is
unstable, whereas the other one (with $y_c=y_c^-$, denoted by
$\phi_1(y)$) is linearly stable. The trivial homogeneous solution
$\phi_0(y)=0$ (denoted by $\phi_0$) exists for any parameter set
and is always linearly stable. These two linearly stable solutions
are the only stable stationary structures under the given albedo
boundary conditions. We will concentrate on the region of values
of $z$, $y_L$ and $k$, where both nonhomogeneous structures exist.

For the system with the albedo b.c. that we are considering here,
the NEP reads \cite{I0}
\begin{equation}
\label{Lyap} {\cal F}[\phi,k,y_L] = \int_{-y_L}^{y_L}\left\{-
\int_0^{\phi(y,t)} \left[ -\phi'+\phi_h \theta(\phi'-\phi_c)
\right] \ d\phi' +{1\over 2} (\partial_y\phi(y,t))^2 \right\} dy +
\left.{k\over 2}\phi(y,t)^2\right|_{\pm y_L}.
\end{equation}
Replacing the explicit forms of the stationary nonhomogeneous
solutions (see for instance Eq.(9) in \cite{SW}), we obtain the
explicit expression \cite{extend2,I1}
\begin{equation}
\label{Lyapunov} {\cal F}^\pm = {\cal F}[\phi_{u,1},k,y_L]
=-\phi_h^2 \, y_c^\pm z + \phi_h^2 \, \sinh(y_c^\pm) \, \frac
{\gamma(k,y_L-y_c^\pm)}{\gamma(k,y_L)},
\end{equation}
while for the homogeneous trivial solution $\phi_0=0$, we have
instead ${\cal F}[\phi_0,k,y_L]={\cal F}^0 = 0$.

Figure 1, part (a) depicts ${\cal F}[\phi,k,y_L]$ as a function of
the system size $y_L$, for a fixed albedo parameter $k$, and a
fixed value of the ratio $\phi_c/\phi_h$ (i.e. fixed value of
$z$). The curves correspond to the nonhomogeneous structures,
${\cal F}^\pm$, whereas the horizontal line stands for ${\cal
F}^0$, the NEP of the trivial solution. We have focused on the
bistable zone, the upper branch being the NEP of the unstable
structure, where ${\cal F}$ attains a maximum, while in the lower
branch (for $\phi = \phi_0$ or $\phi = \phi_1$), the NEP has local
minima. We see that when $y_L$ becomes small, the difference
between the NEP for the states $\phi_u(y)$ and $\phi_1(y)$ reduces
until, for $y_L \approx 0.72$, they coalesce and, for even lower
values of $y_L$, disappear. For completeness and latter use, in
part (b) of Fig. 1 we show ${\cal F}[\phi,k,y_L]$ but now as a
function of $k$, for a fixed value of $y_L$ and the same value of
$z$. Here we see that the initial large difference between the NEP
for the states $\phi_u(y)$ and $\phi_1(y)$ reduces for increasing
$k$ until, for $k \to \infty$, the values for Dirichlet b.c. are
asymptotically reached.

\begin{figure}
\centering
\resizebox{.4\columnwidth}{!}{\rotatebox{-90}{\includegraphics{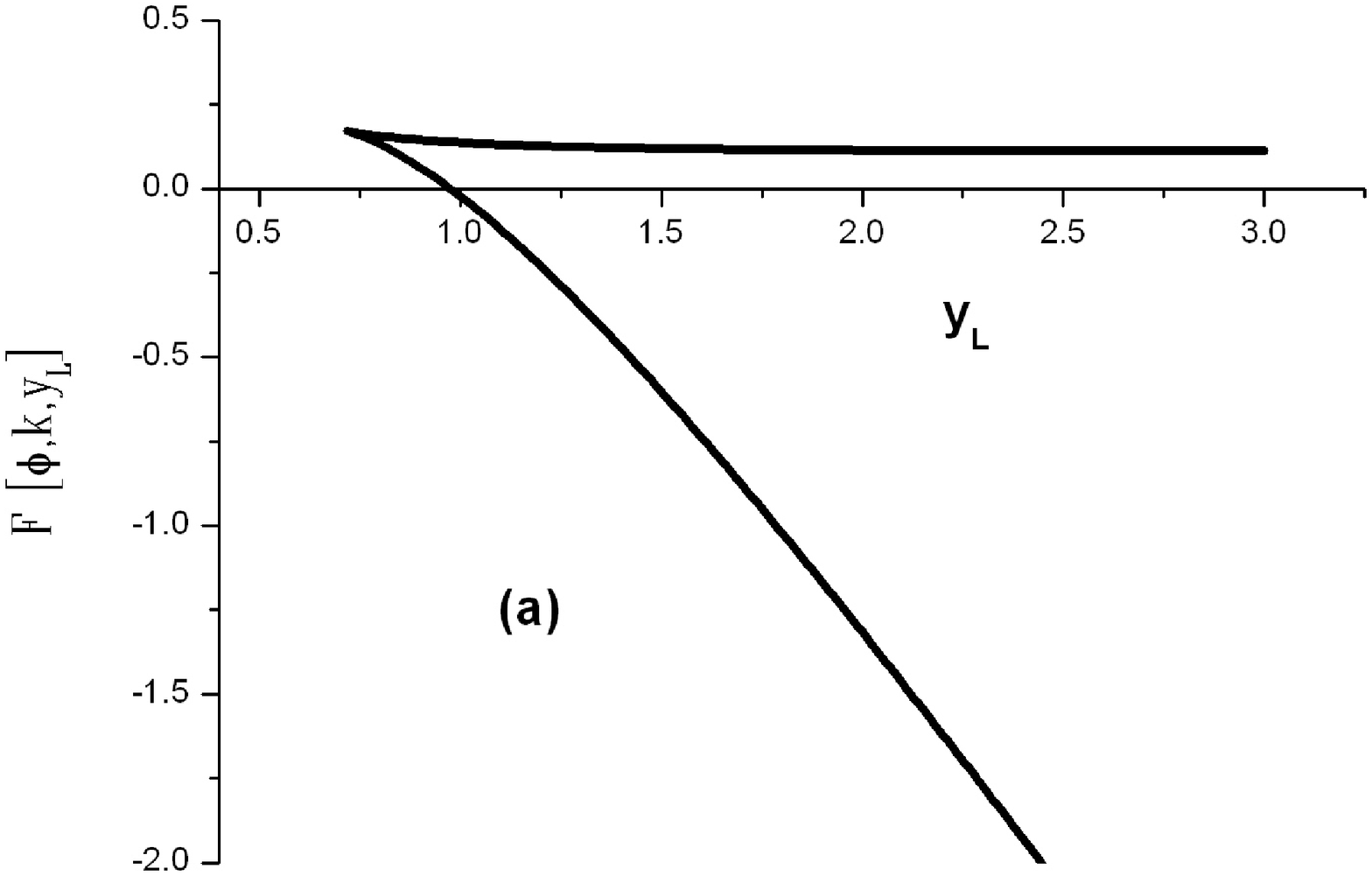}}}
\resizebox{.4\columnwidth}{!}{\rotatebox{-90}{\includegraphics{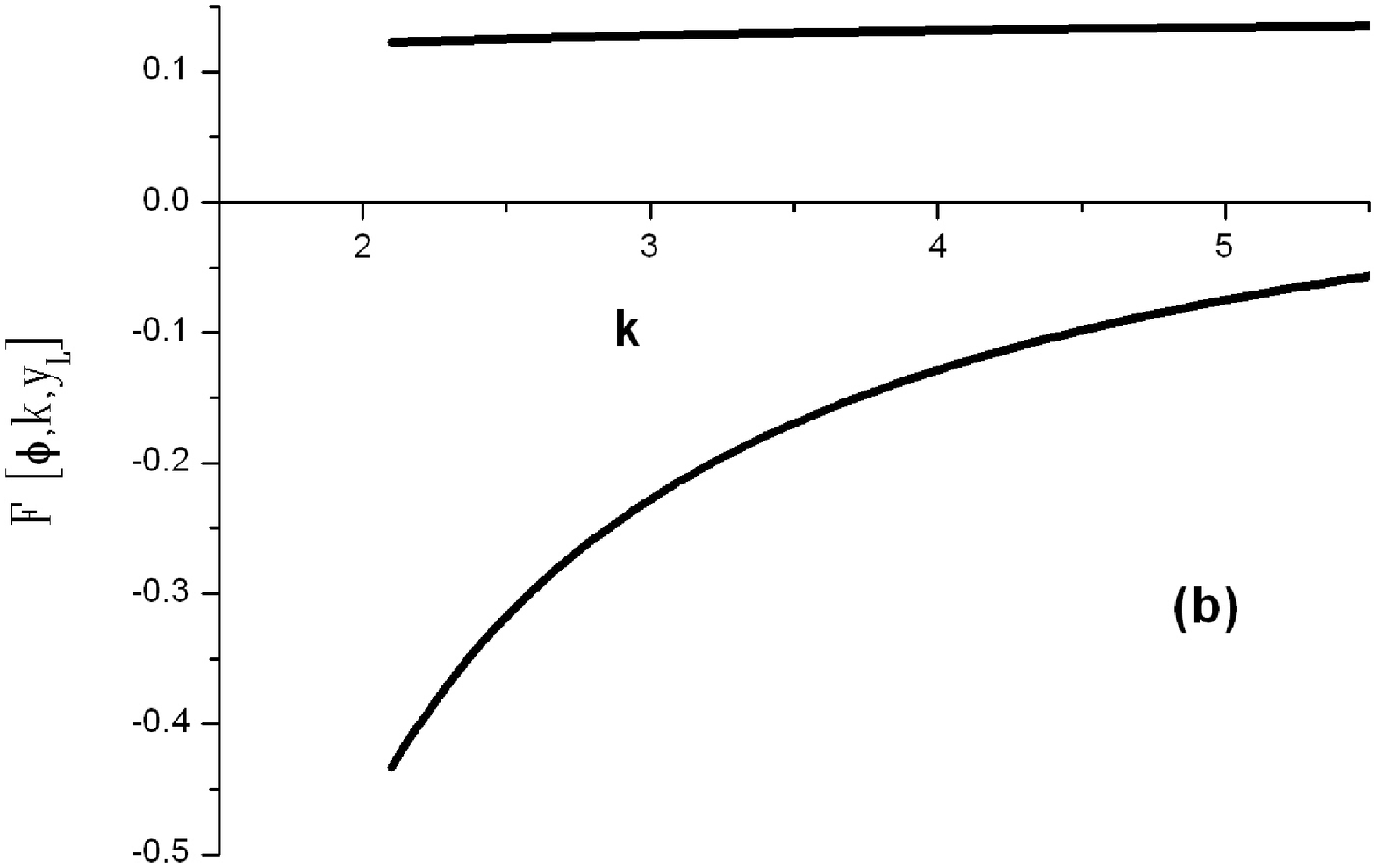}}}
\label{fig1} \caption{NEP evaluated at the stationary solutions
$\phi_0(y)$, $\phi_1(y)$ and $\phi_u(y)$. Part (a)  ${\cal
F}[\phi,k,y_L]$ vs. $y_L$, with $k= 3$. Part (b) ${\cal
F}[\phi,k,y_L]$ vs. $k$, with $y_L = 1.2$. In both cases
$\phi_c/\phi_h = 0.193$.}
\end{figure}

It is important to note that, since the NEP for the unstable
solution $\phi_u$ is always positive and, for the stable
nonhomogeneous structure $\phi_1$, ${\cal F} < 0$ for $y_L$ large
enough, and ${\cal F} > 0$ for small values of $y_L$, the NEP for
this structure vanishes for an intermediate value $y_L = y_L^*$ of
the system size. At that point, the stable nonhomogeneous
structure $\phi_1(y)$ and the trivial solution $\phi_0(y)$
exchange their relative stability.

In order to account for the effect of fluctuations, we include in
the time--evolution equation of our model (Eq.(\ref{Ballast})) a
fluctuation term, that we model as an additive noise source
\cite{extend3b,nsp}, yielding a stochastic partial differential
equation for the random field $\phi(y,t)$
\begin{equation}
\label{noise}
\partial_t \phi(y,t)=\partial^2_{yy} \phi-\phi+\phi_h \theta
(\phi-\phi_c) + \xi(y,t).
\end{equation}
We make the simplest assumptions about the fluctuation term
$\xi(y,t)$, i.e. that it is a Gaussian white noise with zero mean
and a correlation function given by: $\langle\xi(y,t) \,
\xi(y',t')\rangle = 2~\gamma~\delta(t-t')~\delta(y-y')$, where
$\gamma$ denotes the noise strength.

As was discussed in
\cite{extend2,extend2a,extend2b,extend3a,extend3b}, using known
results for activation processes in multidimensional systems
\cite{HG}, we can estimate the activation rate according to the
following Kramers' like result for $\langle\tau\rangle$, the
first-passage-time for the transitions between attractors,
\begin{equation}
\label{tau} \langle\tau _{i}\rangle= \tau_{0} \, \exp \left\{
\frac{ \Delta {\cal F}^{i}[\phi,k]}{ \gamma } \right\},
\end{equation}
where $\Delta {\cal F}^{i}[\phi,k,y_L]={\cal F}[\phi_{u}(y), k,
y_L] - {\cal F}[\phi_{i}(y),k,y_L]$ ($i=0,1$). The pre-factor
$\tau_{0}$ is usually determined by the curvature of ${\cal
F}[\phi,k,y_L]$ at its extreme (minima) and typically is, in one
hand, several orders of magnitude smaller than the average time
$\langle\tau\rangle$, while on the other does not change
significatively when changing the system's parameters. Hence, in
order to simplify the analysis, we assume here that $\tau_{0}$ is
constant and scale it out of our results. The behavior of
$\langle\tau\rangle$ as a function of the different parameters
($k$, $\phi_{c}$) was shown in
\cite{extend2,extend2a,extend2b,extend3a,extend3b,I0,I1}.

We assume now that the system is subject (adiabatically) to an
external harmonic variation of the parameter $\phi_{c}$:
$\phi_{c}(t) = \phi_{c} + \delta \phi_{c} \cos(\omega t)$
\cite{extend2a,extend3b}, and exploit the ``two-state
approximation" \cite{RMP} as in
\cite{extend2a,extend2b,extend3a,extend3b}. For all details on the
general two-state approximation we refer to \cite{extend3a}.

Up to first-order in the amplitude $\delta \phi_{c}$ (assumed to
be small in order to have a sub-threshold periodic input) the
transition rates $W_i$ take the form
\begin{equation} W_{i} =
\tau_0^{-1} \exp \left\{ - \frac{ \Delta {\cal F}^{i}
[\phi,k,y_L]}{ \gamma } \right\}
\end{equation}
where
\begin{equation}
\label{modul} \Delta {\cal F}^{i}[\phi,k,y_L]=\Delta {\cal
F}^{i}[\phi,k,y_L] + \delta \phi_{c} \Bigl( \frac{\partial \Delta
{\cal F}^{i}[\phi,k,y_L]}{\partial \phi_{c}}
\Bigr)_{\phi_{c}=\phi_{c}^*} \cos (\omega t).
\end{equation}
This yields for the transition probabilities
\begin{equation}
\label{www} W_{i} \simeq \frac{1}{2} \Bigl(\mu_{i} \mp \alpha_{i}
\frac{\delta \phi_{c}}{\gamma} \cos(\omega t) \Bigr),
\end{equation}
where $\mu_{i} \approx \exp(-\Delta{\cal F}^{i} [\phi,k,y_L])$ and
$\alpha_{i} \approx \pm \mu_{i} \frac{d\Delta{\cal F}^{i}}
{d\phi_{c}}|_{\phi_{c}}$ ($i=1,2$). Using Eq. (\ref{Lyapunov}),
$\frac{d\Delta{\cal F}^{i}}{d\phi_{c}}|_{\phi_{c}^*}$ can be
obtained analytically.

These results allows us to calculate the autocorrelation function,
the power spectrum and finally the SNR, that we indicate by $R$.
The details of the calculation were shown in \cite{extend3a} and
will not be repeated here. For $R$, and up to the relevant
(second) order in the signal amplitude $\delta \phi_{c}$, we
obtain \cite{extend3a}
\begin{equation}
\label{snr} R= \, \frac{\pi}{4\,  \mu_1 \, \mu_2} \frac{(\alpha_2
\, \mu_1 + \alpha_1 \, \mu_2)^2}{\mu_1 + \mu_2}.
\end{equation}
We have now all the elements required to analyze the problem of
SSSR.

Figure 2 shows the typical behavior of SR, but now --in the
horizontal axis-- the noise intensity is replaced by the the
system length $y_L$, for fixed values of $k$, $\gamma$ (the noise
intensity) and $\phi_c/\phi_h$. Such a response is the expected
one for a system exhibiting SSSR. Within the context of NEP, it
results clear that, in this kind of systems, the phenomenon arises
due to the breaking of the NEP's potential symmetry. That is: both
attractors change their relative stability due to the variation of
$y_L$ as shown in Fig. 1.a. Hence, within this framework, SSSR
arises as a particular case of the general discussion in
\cite{extend3a}.

\begin{figure}
\centering
\includegraphics[width=7cm,angle=-90]{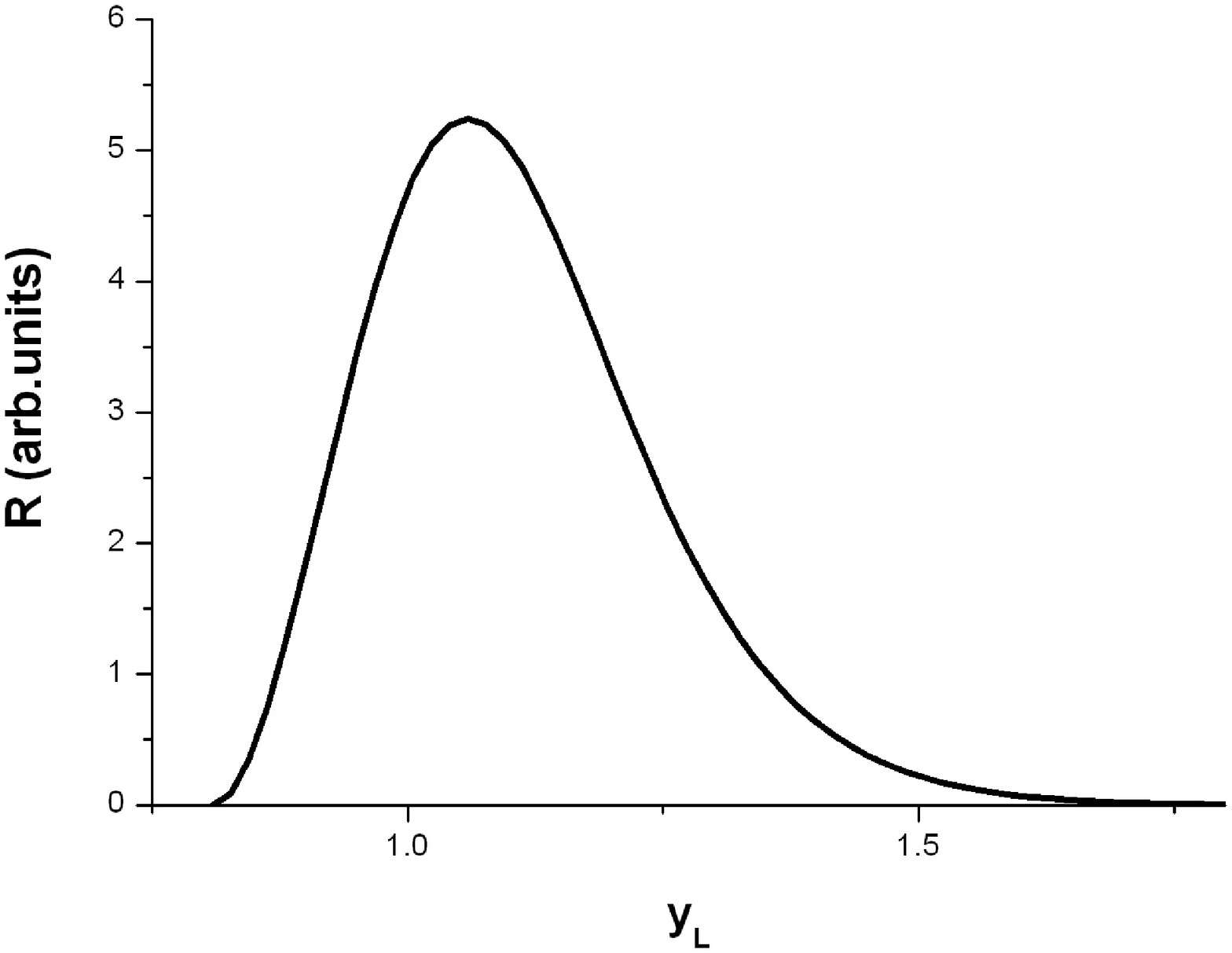}
\label{fig2} \caption{SNR, vs. $y_L$, for $k = 3.$, $\gamma = 0.1$
and $\phi_c/\phi_h = 0.193$.}
\end{figure}

Let us now change the point of view. In Fig. 3 we show the curves
of the SNR as a function of $k$, while keeping fixed values of
$y_L$, and $z$. When $k$ is not too large, indicating a high
degree of reflectiveness at the boundary (or a reduced exchange
with the environment), we see that the SNR changes for $k$ varying
from low to larger values. Remember that a large value of $k$
indicates that the system boundaries become absorbent. Also, the
{\em robustness} of the systems' response when changing $k$, a
parameter that somehow indicates the degree of coupling with the
environment, is apparent. According to the previous argument
--about the breaking of NEP's symmetry-- from Fig. 1.b this is the
expected result.

\begin{figure}
\centering
\resizebox{.6\columnwidth}{!}{\includegraphics[angle=90]{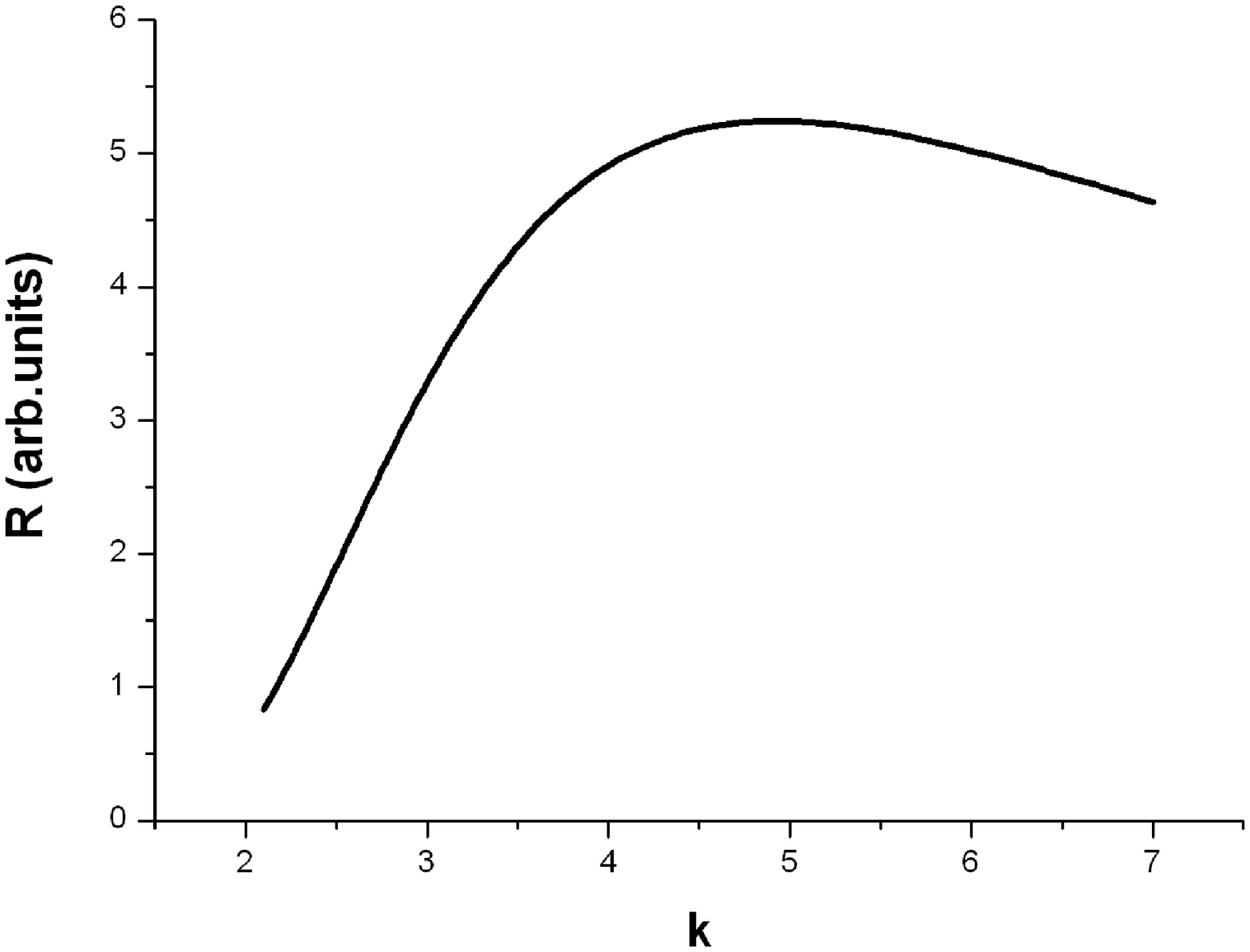}}
\label{fig3} \caption{SNR, vs. $k$ for $y_L = 1.2$, $\gamma = 0.1$
and $\phi_c/\phi_h = 0.193$.}
\end{figure}

As a final remark, let us consider one of the models discussed in
\cite{SSSR4} from the point of view of the above indicated
approach. The model we refer to is described by a set of coupled
nonlinear bistable oscillators
\begin{eqnarray}
\label{PIKOV1} \dot{x}_{j} & = & x_{j} - x_{j}^3 +
\frac{\varepsilon}{N} \sum_{k=1}^{N} (x_{k}-x_{j}) +
\sqrt{2\,\gamma}\, \xi_{j}(t) + f_j(t),\nonumber \\
\dot{x}_{j} & = & - \frac{\partial}{\partial \,x_{j}} U(\{ x \},t)
+ \sqrt{2\,\gamma}\, \xi_{j}(t),
\end{eqnarray}
with $f_j(t) = A \cos (\omega t)$, $\{ x \} = (x_{1},x_{2},..
,x_{N})$, and
\begin{eqnarray}
\label{PIKOV2} U(\{ x \},t) & = & \sum_{j=1}^{N} u_0(x_j) - A
\cos (\omega t) \sum_{j=1}^{N} x_{j} \nonumber \\
& = & U_0(\{ x \}) - A \cos (\omega t) \sum_{j=1}^{N} x_{j} \nonumber \\
& = & \sum_{j=1}^{N} \left( \frac{x_{j}^4}{4} - \frac{x_{j}^2}{2}
\right) + \frac{\varepsilon}{2 \, N} \sum_{j=1}^{N} \sum_{k=1}^{N}
(x_{k}-x_{j})^{2} - A \cos (\omega t) \sum_{j=1}^{N} x_{j}.
\end{eqnarray}
Due to the structure of Eq. (\ref{PIKOV1}) it is clear that the
potential function in Eq. (\ref{PIKOV2}) is the NEP of this
problem and --for $A=0$-- that the stationary distribution of the
multidimensional Fokker-Planck equation associated to Eq.
(\ref{PIKOV1}) is
\begin{equation}
\label{PIKOV3} P_{stat}(\{ x \}) \approx \exp \left(- \frac{U_0(\{
x \})}{\gamma} \right).
\end{equation}
Clearly, this potential has two attractors corresponding to
$x_{1}=x_{2}=... =x_{N}=\pm 1$, and a barrier separating them at
$x_{1}=x_{2}=... =x_{N}=0$.

Now, exploiting the same scheme as before, but reduced to the
symmetric case (as both attractors have the same ``energy"), we
get
\begin{equation}
\label{PIKOV4} SNR \approx \exp \left(- \frac{\triangle U_0(\{ x
\})}{\gamma} \right) \approx \frac{N}{\gamma} \, \exp \left(-
\frac{N\, \triangle u_0(X)}{\gamma}\right),
\end{equation}
where $X$ is a ``collective coordinate" (that can be approximately
interpreted as $X \approx \frac{1}{N}\sum_{j=1}^{N} x_{j}$), and
$\triangle u_0(X) = u_0(x = \pm 1) - u_0(x = 0)$. This SNR clearly
shows similar SSSR characteristics as those described in
\cite{SSSR4}. However, as in this situation the NEP's symmetry is
retained when varying $N$, and we could speak of a {\em genuine}
SSSR.

The above results clearly show that the ``nonequilibrium
potential", even if not known in detail \cite{quasi}, offers a
very adequate framework to analyze a wide spectrum of noise
induced phenomena in spatially extended or coupled systems. Within
this framework the phenomenon of SSSR looks, as in other aspects
of SR in extended systems \cite{extend3a}, as a natural
consequence of the breaking of the symmetry of the NEP. In
addition, we have seen that through the variation of its coupling
with the surroundings, a system can increase the robustness of its
response to an external signal, opening new possibilities of
analyzing and interpreting the behavior of biological systems.

An important conclusion to be drawn from the identification of the
Lyapunov functional with a ``thermodynamical-like potential", is
that for a wide range of parameters where the system is
essentially bistable (with both attractors not necessarily having
the same ``energy"), the problem admits a one--dimensional
analogue \cite{I0,I1,extend3b}. This feature is in contrast with
the infinite dimensional character of the whole function space,
and has been exploited to strongly simplify the analysis. \\

{\bf ACKNOWLEDGEMENTS:} The author thanks R. Toral for calling his
attention to this problem, and C. Tessone and M.A. Rodriguez for
fruitful discussions. He also acknowledges partial support from
ANPCyT, Argentine, and thanks the European Commission for the
award of a {\it Marie Curie Chair}.


\end{document}